# Ionization-density-dependent scintillation pulse shape and mechanism of luminescence quenching in LaBr$_3$:Ce

Jirong Cang[1,2], XinChao Fang[1,2], Zhi Zeng[1,2], Ming Zeng[1,2,*], Yinong Liu[1,2], Zhigang Sun[3], Ziyun Chen[4]

[1] *Key Laboratory of Particle & Radiation Imaging (Tsinghua University), Ministry of Education, China*
[2] *Department of Engineering Physics, Tsinghua University, Beijing, 100084, China*
[3] *Institute of Materials Science & Chemical Engineering, Ningbo University, Ningbo, 315211, China*
[4] *School of Electric Information & Electrical Engineering, Shanghai Jiao Tong University, Shanghai, 200240, China*

Pulse shape discrimination (PSD) is usually achieved using the different fast and slow decay components of inorganic scintillators, such as BaF$_2$, CsI:Tl, etc. However, LaBr$_3$:Ce is considered to not possess different components at room temperature, but has been proved to have the capability of discriminating gamma and alpha events using fast digitizers. In this paper, ionization density-dependent transport and rate equations are used to quantitatively model the competing processes in a particle track. With one parameter set, the model reproduces the non-proportionality response of electrons or alpha particles, and well explains the measured α/γ pulse shape difference. In particular, the nonlinear quenching of excited dopant ions, Ce$^{3+}$, is confirmed herein to mainly contribute observable ionization α/γ pulse shape differences. Further study of the luminescence quenching can also help to better understand the fundamental physics of nonlinear quenching and thus improve the crystal engineering. Moreover, based on the mechanism of dopant quenching, the ionization density-dependent pulse shape differences in other fast single-decay-component inorganic scintillators, such as LYSO and CeBr$_3$, are also predicted and verified with experiments.

## I. INTRODUCTION

Inorganic scintillators have been widely used to monitor γ/X-rays or particle beams in applications of medical imaging and dosimetry, security check, high-energy physics and the hightech industry. One may hope to discover and/or engineer a scintillator with both a high light yield, high energy resolution, small non-proportionality and fast rise and decay times. Non-proportionality is departure from proportional response of the number of scintillation photons produced versus energy of the primary particle, which has long been considered as a fundamental limitation of the intrinsic energy resolution of scintillators[1-6]. Great efforts in both experimental measurements[7-10] and theoretical models[2,11-18] have been made to study the non-proportionality of inorganic scintillators. It is generally considered that non-proportionality arises from the variation in excitation densities depending partly on the particle energy. It is thought that high excitation densities lead to strong particle-particle interactions and nonlinear quenching and that, at low excitation densities, slow thermalization can lead to extensive charge separation[12,13,18]. Moreover, the ionization density-dependent scintillation processes can not only lead to the non-proportionality of the scintillation yield to incident particle energy, but also the differences of pulse shapes to incident particle energy, e.g CsI:Tl[19], or different types of particles, e.g n/γ discrimination. It looks like the non-proportionality and pulse shape differences would be connected at the level of nonlinear interactions and details of transport, which may provide a basis for using the underlying extra information in pulse shape to correct the non-proportionality and thus improve the energy resolution[20].

The best scintillation performance in terms of light yield (70000 Photons/MeV), decay time (~16 ns), and energy resolution (~2.6% at 662 keV) has been achieved for $LaBr_3$:Ce. In recent years, many research groups, including us, have studied the α/γ pulse shape discrimination (PSD) in $LaBr_3$:Ce crystals, but contradictory results were achieved[21-28]. It is usually considered that scintillators with multicomponent decay curves can have the ability of pulse shape discrimination (PSD) between gamma rays and massive (for example, proton and alpha) particles. However, $LaBr_3$:Ce is considered to only possess single decay component, but has been proved to have the capability of discriminating gamma and alpha events using fast digitizers at room temperature, which can be used to reduce the intrinsic alpha background from $^{227}$Ac contamination and improve its application for low-activity gamma ray spectroscopy. The physical mechanism of such PSD capability of $LaBr_3$:Ce was remained unclear. Such pulse shape difference, as well as strong quenching of alpha particles, in $LaBr_3$:Ce was generally explained to be the exciton-exciton annihilation (bi-molecular decay)[29,30]. However, the precisely measured rise time, associated with self-trapped excitons, of $LaBr_3$:5%Ce pulse shape is as fast as approximately 1 ns[31], which seems not enough to cause such small but significant difference. A complete and quantitative model illustrating the scintillation mechanism of $LaBr_3$:Ce is still needed.

The model in this paper is based on the coupled rate and transfer equations proposed by Lu, *et al.* [13,14], which include most of the already known scintillation processes and has been applied to CsI:Tl crystals to explain the non-proportionality and energy-dependent pulse shapes of gamma rays. Regarding the classification of inorganic scintillators, $LaBr_3$:Ce belongs to the multivalent halide[32] and shows different characters from alkali halide, such as CsI, NaI. Coupled rate and transport equations describing the main physical processes of $LaBr_3$:Ce crystal, at room temperature, are established. The model computes the evolution of excitations (including electrons, holes, excitons, and excited dopant ions) over time and space in electron tracks by solving coupled rate and transport equations describing both the movement and the linear and nonlinear interactions of the excitations along the ionization track. The tracks are initially very narrow before hot and thermalized carrier diffusion takes effect. A cylindrical Gaussian radial profile with a track radius of $r_0(\sim 3 \text{ nm})$ is generally adopted to describe the distributions of excitations on a series of small cells along the ionization track as reviewed in [33]. The initial ionization density values vary from cell to cell along the track with the variation in linear energy deposition rate dE/dx and the model calculate the local light yield, of which the weighted average is the light yield, for each local value of dE/dx. Besides the light yield output, the evolution of excitations (emission intensity) over time is also responsible for the pulse shapes, which allows us to validate the model more thoroughly. Furthermore, alpha particle is also an alternative way to reach even higher density of excitation than that of electrons, which can be used to study and validate the ionization density-dependent physical processes even further. The study of scintillation mechanism of inorganic scintillators will not only help to understand the characteristics of scintillators, but also guide the optimization of new scintillator materials and the development of new radiation detection methods.

## II. THE MODEL AND ITS PARAMETERS IN PARTICLE TRACKS
### A. The scintillation mechanism in LaBr$_3$:Ce

Many studies about the properties of LaBr$_3$:Ce have been carried out and form the basis of the scintillation mechanism. X-ray excited emission spectra of LaBr$_3$:Ce at 125 K shows that both STE broad band and double peaked 5d→4f cerium emission are observed. The contribution of the STE broad band decreases from 70% to 37% to 8% for a Ce concentration of 0.2%, 0.5%, and 5%, respectively, which shows an anti-relation between the STE emission and Ce concentration. Moreover, the emission spectra for LaBr$_3$:5%Ce, at room temperature, only arises from the double peaked 5d→4f cerium emission[34]. The scintillation time profiles measured with delayed coincidence method or time-correlated single photon counting (TCSPC) method reveal more information about the scintillation mechanism of LaBr$_3$:Ce. Though STEs do not contribute the emission directly, at room temperature, according to the emission spectra of LaBr$_3$:Ce, Glodo *et al.*[35] suggested a diffusion and energy transfer model of STEs to Ce based on the linear relation between the logarithms of Ce concentration and rise time of scintillation time profiles[35]. Later on, Bizarri and Dorenbos proposed a STE-transport based model to account for gamma excited luminescence in LaBr$_3$ with different Ce concentrations and at different temperatures[36]. In that model, three processes are proposed, i.e. (i) Process I, the prompt sequential capture of electron and holes by Ce. (ii) fast process II, thermally activated energy transfer from self-trapped excitons situated in the close surrounding of a cerium ion to that cerium ion. (iii) slow process II, thermally activated migration of STEs over a distance to encounter a Ce dopant followed by energy transfer from STE to Ce. However, in terms of the fitting of the scintillation pulse shape, we want to argue that the decay time, which arises only from 5d→4f cerium emission and is insensitive to temperature[37], should be fixed as ~16 ns to fit the decay curves and then to discuss the model. Recently, the study of picosecond absorption spectroscopy by Li *et al.*[38] shows that $Ce^{3+}$ in lanthanum bromide is apparently not a good electron or hole trapper. At least self-trapped holes in $LaBr_3$ appear to be better at capturing electrons than $Ce^{3+}$ ions are at capturing holes. The energy transport from host to activator is responsible for the scintillation of $LaBr_3:Ce^{3+}$ proceeds by STE creation within 1 ps and then energy transfer more than by the sequential trapping of holes and electrons on $Ce^{3+}$ ions [38]. Detailed scintillation rise-time measurements in $LaBr_3$:Ce with fast coincidence methods by Glodo *et al.* [35] and Seifert *et al.* [31] have identified a fast stage and a slower stage of scintillation rise for $LaBr_3$:5%Ce given as 380 ps[35] or 270 ps[31], and 2.2 ns[35] or 2.0ns[31], respectively. Furthermore, the ~300 ps process is further suggested by Li *et al.*[38] as the dipole-dipole transfer to Ce from STEs created in the close neighborhood, and the ~2.1 ns process is the thermally activated migration and energy transfer at room temperature in 5%Ce-doped $LaBr_3$. For $LaBr_3$:Ce with higher concentration, it is possible that Ce ions can be excited directly. For example, the extreme case with 100% Ce concentration, $CeBr_3$ should be the direct excitation and radiative emission of $Ce^{3+}$ ions. To conclude, the scintillation mechanism in $LaBr_3$:Ce with lower concentration, for example less than 5%, at room temperature is mainly first the creation of host STEs, then the migration and/or energy transfer from STEs to $Ce^{3+}$ ions and finally the radiation emission of excited $Ce^{3+*}$ ions:

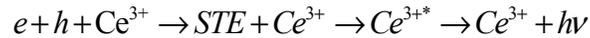
$$e + h + Ce^{3+} \rightarrow STE + Ce^{3+} \rightarrow Ce^{3+*} \rightarrow Ce^{3+} + h\nu$$

A model of excitation transport and interaction in a particle track is established to fully explore the scintillation mechanism of $LaBr_3$:Ce, shown as Fig. 1. In detail, the model can be separated into two main stages. The first stage includes the hot and thermalized free charge carrier diffusion, electric-field transport and the form of STEs. The second stage includes the migration and energy transfer from STEs to $Ce^{3+}$ ions, second-order nonlinear quenching between STEs or between excited $Ce^{3+}$ ions, and radiative recombination. In particular, apart from the direct excitation of $Ce^{3+}$ ions, only STEs formed in the first stage can have access to the second stage and transfer their energy to Ce and account for scintillation.

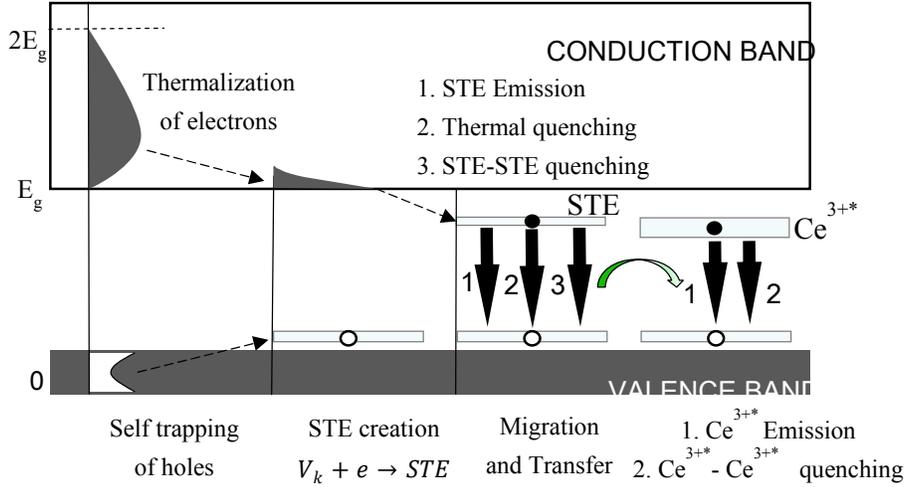

Fig. 1. Illustration of scintillation processes of physical mechanism in LaBr$_3$:Ce$^{3+}$ crystals, including the conversion of free electrons and hole, the thermalization of hot electrons and instant self trapping of holes, the charge carrier diffusion and the formation of STE, the migration and energy transfer from STE to Ce ions, and finally luminescence of excited Ce$^{3+*}$ ions.

## B. Diffusion-limited rate and transport equations for local light yield and decay curve versus on-axis initial excitation densities in particle tracks

The diffusion-limited rate and transport equation illustrating the local light yield (LY) and scintillation decay curve for LaBr$_3$:Ce in our model are calculated using Eqs. (1)-(5).

$$\frac{dn_e}{dt} = G_e + D_e \nabla^2 n_e + \mu_e \nabla \cdot n_e \vec{E} - K_{1e} n_e - B n_e n_h - K_3 n_e n_h n_e - K_3 n_e n_h n_h \quad (1)$$

$$\frac{dn_h}{dt} = G_h + D_h \nabla^2 n_h + \mu_h \nabla \cdot n_h \vec{E} - K_{1h} n_h - B n_e n_h - K_3 n_e n_h n_e - K_3 n_e n_h n_h \quad (2)$$

$$\frac{dN_{STE}}{dt} = G_{STE} + D_{STE} \nabla^2 N_{STE} - \left(R_{STE} + Q_{STE} + S_T\right) N_{STE} + B n_e n_h - K_{2E}(t) N_{STE}^2 \quad (3)$$

$$\frac{dN_{Ce^*}}{dt} = G_{Ce^*} + S_T N_{STE} - R_{Ce^*} N_{Ce^*} - K_{2Ce^*}(t) N_{Ce^*}^2 \quad (4)$$

$$S_T = S_T^0 \cdot \frac{n_{Ce}^0 - N_{Ce^*}}{n_{Ce}^0} \quad (5)$$

Eqs. (1) and (2) describe the first stage of free electrons and holes. The first stage also consists of two processes. The hot charge carriers, especially hot electrons, first diffuse outward and get thermalized. After thermalization, the distinct hot diffusion of electrons and holes will create an inner electric filed. Out-diffused electrons will then be driven back to form STEs. Eqs. (3) -(5) describe the second stage of STEs and $Ce^{3+}$ ions. We will now describe each of the terms in these equations from the view of scintillation processes and their significant influences on the light yield and scintillation time profiles.

During the thermalization process of first stage, which lasts for several picoseconds, the electrons and holes are excited with high excitation density and therefore will diffuse outward and lose their excess kinetic energy (~eV) by impacting with the crystal lattice. Meanwhile, the outward free electrons and holes may suffer from linear trapping of defects or the third Auger quenching process, which may have a significant effect on the light yield of the scintillators and is ionization density-dependent. During the thermalization, part of the free electrons and holes pair together and form STEs within several picoseconds, known as instant STEs, while the others remain free. For simplification, the formation process of STEs within the thermalization process, which is approximately 1 ps for LaBr$_3$[10,31,38], is regarded as the instant generation term.

The first terms of Eqs. (1)-(4), $G_{e,h}$, $G_{STE}$, and $G_{Ce^*}$, are respectively the generation terms of free electrons and holes, instant STEs, and directly excited $Ce^{3+}$ ions immediately after thermalization. Generally, a cylindrical Gaussian radial profile with a track radius of $r_0$ (~3 nm) is adopted to describe the distributions of excitations (including electrons, holes, excitons, and excited dopant ions) as reviewed in [33]. The radial distribution and its magnitude, the on-axis excitation density $n_0$, is shown as follows:

$$n(r, t=0) = n_0 \exp\left(-r^2/r_0^2\right) \tag{6}$$

$$n_0 = \frac{dE/dx}{\pi r_0^2 \beta E_{gap}} \tag{7}$$

where dE/dx is the energy-dependent linear energy transfer of a particle (electron/α). $\beta E_{gap}$ is the average energy invested per electron hole pair. $r_0$ is the initial track radius, which is determined by the diffusion of hot holes during the thermalization process. In particular, the ionization densities produced on the end of an electron track can be as high as ~$10^{20}$ cm$^{-3}$ and range out in a very short distance.

The second terms $D_{e,h}$ and $D_{STE}$ of Eqs (1)-(3) denote the diffusion of free electrons and holes and self-trapped excitons (STEs). As described in [13], holes in alkali halides are self-trapped very quickly, which is 50 fs for NaI from quantum molecular dynamics calculations. Due to such rapid self-trapping, the hole equation (2) is simply written in terms of the density of the self-trapped holes (STHs), $n_h$, diffusing with the hopping diffusion coefficient of the self-trapped holes. However, the thermalization of electrons is somewhat complicated, and the cooling of the hot electrons in LaBr$_3$:Ce is rather slow due to its low longitudinal optical phonon frequency ($\omega_{LO}$)[10]. A mean thermalization time of 1 ps is characterized for LaBr$_3$ due to its very similar $\omega_{LO}$ to NaI crystal[10,39], of which the mean thermalization time calculated by Wang *et al.*[40] is approximately 1 ps using the NWEGRIM Monte Carlo code at PNNL[40]. The longitudinal optical phonon frequency is $3.6\times10^{13}$ s$^{-1}$ and $3.47\times10^{13}$ s$^{-1}$ for LaBr$_3$ and NaI, respectively. The thermalization time was further supported by the picosecond absorption spectroscopy experiment to be less than 1 ps under two-photon excitation of the host producing carriers near the band edges[38]. According to Wang *et al.*[40], hot electrons run outward to a radial peak of approximately 30 nm for fluorides (CaF$_2$ and BaF$_2$) and 50 nm for iodides (NaI and CsI ), with a tail extending as far as 100 to 200 nm. Since $D_e$ is a function of the electron temperature $T_e$, it is difficult to precisely model the hot diffusion process. A similar step-wise time-dependent electron diffusion coefficient proposed by Lu, *et al.* [13,14] was adopted such that $D_e(t<\tau_{hot})$ has a constant value to reproduce the distribution of electrons with peaks $r_{hot}$ at several tens of nanometers. However, based on the first principle calculations[18,39] and the "decision tree" of inorganic scintillators proposed by Li *et al.*[32], LaBr$_3$:Ce, unlike alkali halide, is a multivalent compound with a dense and flat set of 4f conduction bands approximately 3.5 eV above the conduction band minimum (cbm) and thus smaller electron group velocities for a smaller electron distribution range. The fourth terms, $K_{1e}$ and $K_{1h}$, of Eqs. (1) and (2) are the trapping from deep traps or defects of the crystal. The sixth and seventh terms, $K_3$, of Eqs. (1) and (2) are the third-order Auger recombination rates of free carriers. Since the hot diffusion term of free electrons will greatly reduce the ionization density, the Auger quenching process in LaBr$_3$:Ce is trivial based on the parameters measured by a laser Z scan of similar materials, which will be disscussed in Sec.II C.

After thermalization, the excitations will diffuse with the thermalized diffusion rate, which is described by the Einstein equation, $D=\mu kT$. In particular, the diffusion rate of electrons after thermalization is calculated by $D_e(t>\tau_{hot})=\mu_e kT$. Due to the significant differences in the diffusion rates of hot electrons and self-trapped holes (STHs or V$_k$ centre), a distinct spatial distribution of the electrons and holes will create an electric field. The migration of the free thermalized electrons and holes will proceed due to the electrostatic forces. The direction of the electron current reverses from outward to inward as the thermalized conduction electrons are collected back toward the line charge STHs where recombination can occur. The third terms, $\mu_e$ and $\mu_h$, of Eqs. (1) and (2) represent the electric field driven currents. The fourth term, $Bn_en_h$, in Eqs. (1) and (2) is the bimolecular exciton formation characterized by rate constant $B$ and proportional to the product of the electron and hole densities at a given location and time. The exciton formation term, $-Bn_e n_h$, is a loss term in Eqs. (1) and (2) but the extra source term in Eq. (3)'s exciton density of the STEs apart from the initial creation of the instant STEs during the thermalization process of the first stage.

As illustrated before, the processes of the sequential capture of free electrons or holes by dopant ions are almost absent, since $Ce^{3+}$ is not a good trapper for both electrons and holes in LaBr3:Ce. The main part of the transport process is the migration and energy transfer from the STEs to the $Ce^{3+}$ centres. In the third term of Eq. (3), $R_{STE}$ and $Q_{STE}$ are the rates of the radiative decay and thermal quenching of the STEs. $S_T$ is the energy transfer rate of the STEs to $Ce^{3+}$. Since the energy transfer rate is proportional to the density of the unexcited $Ce^{3+}$ ions, Eq. (5) denoting such a relationship is also introduced. Compared with the model proposed by Bizarri and Dorenbos[36], a second-order dipole-dipole quenching process, $K_{2E}(t)$, between the excited STEs is introduced, which plays an important role in situations of high excitation densities and account for a lower light yield and possibly different decay curves. Dipole-dipole annihilation is a case of Förster transfer from one excited dipole to another excited dipole rather than to a ground-state dipole, resulting in the annihilation of the first dipole and possibly ionisation of the doubly excited second dipole. The second-order rate constant can be expressed for immobile species as:

$$K_2(t) = \frac{2}{3}\pi^{3/2} R_{dd}^3 \left(t\tau_R\right)^{-1/2}, \qquad (8)$$

where $\tau_R$ is the radiative lifetime of the excited state and $R_{dd}$ is the Förster transfer radius depending on the overlap of the emission and absorption bands.

The first term, $G_{Ce^*}$, of Eq. (4) is the directly creation of the excited $Ce^{3+}$ states. Only electrons with enough excess kinetic energy to excite the $Ce^{3+}$ ions can create such initial excited $Ce^{3+}$ states, which is especially significant for crystals with high dopant concentrations. The second term, $S_T N_{STE}$, of Eq. (4) is the main source of the excited $Ce^{3+*}$ states from the energy transfer of the STEs. The third term, $-R_{Ce^*} N_{Ce^*}$, of Eq. (4) is the radiative recombination of the excited $Ce^{3+*}$ states and is the dominant source of luminescence. In particular, the second dipole-dipole quenching between excited $Ce^{3+*}$ ions, $K_{2Ce^*}(t)$, is proposed and verified to account for the non-proportionality and pulse shape difference for high excitations, which will be discussed later in Sec.II C.

### C. The material input parameters

This section details the parameters, listed in Table I, used in our model. Most of the parameters were found directly in the literature when possible, or scaled by quantitative physical arguments from parameters known in similar materials.

The initial gaussian ionization radius of the track, $r_0$, is normally considered as 3 nm, which was evaluated by Z scan[10] and kinetic Monto Carlo simulations in NaI and CsI[16,18,40]. LaBr$_3$ also belongs heavier halide and the holes are self-trapped[41] as in the case of NaI and CsI according to the simulation, thus the same parameter $r_0$=3 nm was used for LaBr$_3$.

The value of $\beta E_{gap}$=13 eV, average energy invested per electron-hole pair, was adopted from the best-known light yield, 77000 photons/MeV, of LaBr$_3$:Ce,Sr. The co-doping of 100 ppm Sr enhances the shallow trapping of the free electrons from the capture of the defects and thus increases the light yield. Such small co-doped Sr should not change the number of ionized electron/hole pairs, which is determined by the host LaBr$_3$ and heavy-dopant Ce ions. $\beta$ is calculated as 2.2 based on the band gap of 5.9 eV. LaBr$_3$ is an outstanding inorganic scintillator with high light yield, a value of 2.2 is reasonable compared with a value of 2.5 in most materials.

The thermalized electron mobility $\mu_e$ in LaBr$_3$:Ce is dependent on the dopant Ce concentration and can be calculated based on the ionized impurity scattering[42]. The thermalized conduction electron diffusion coefficient $D_e$ is given in terms of $\mu_e$ by the Einstein relation, $D = \mu k T / e$.

The mobility of self-trapped holes can be estimated from their thermal hopping rate[40,43]. Since there are no direct measurements of this rate in LaBr$_3$, parameter from CsI crystal is taken as a reference since they all belongs to heavier halide. Moreover, the mobility of self-trapped holes is very small, which is not a sensitive parameter in the model, and can be even set to 0 as approximation. The diffusion coefficient, which can calculated with Einstein relation, of STE is considered the same as STH in our model, since recent electronic structure calculations have provided evidence that they are equally mobile in NaI [44,45].

The thermalization process is hypothesized that greatly affects the non-proportionality of the gamma or electron response[13,32]. The fraction of free cairers, $\eta_{e,h}$, and the thermalization distance, $r_{hot}$, are the two significant parameters that affect the non-proportionality. Williams and co-workers hypothesize that extensive charge separation, during the thermalization process, combined with the following electrostatic attraction is the root cause for non-proportionality at high incident energies[13,32]. Later on, Prange et al.[18] (PNNL) simulated the electron thermalization of six scintillating crystals using microscopic kinetic Monte Carlo model and further supported the hypothesis proposed by Williams. They proposed that the thermalization distance are positively correlated with measured non-proportionality. Whereas, the model proposed by Payne et al. [2,11,46] also showed that the fraction of the excitations that are free carriers significantly influences the non-proportionality. The Payne-Onsager model is based on theories by Onsager, Birks, Bethe-Bloch, Landau and by appropriate choice of parameters and can well reproduce the experimental data[2,11]. For a given dE/dx, the Birks and Onsager mechanisms are the two main processes compete each other. Actually, the Onsager mechanism arises from Coulombic attraction of carriers, which is similar as the process of charge separation proposed by Williams. Based on the fitted parameters of Payne-Onsager model, $\eta_{e,h}$ is one of the key parameters to cluster the scintillators. Alkli halides tend to possess a high fraction of free carriers and show more "hump" size in electron/gamma response curves. The fitted $\eta_{e,h}$ parameters with Payne-Onsager model are also in good accordance with the kinetic Monto Carlo simulation results of Wang *et al* (2012, PNNL) in CsI(Tl) and NaI(Tl) crystals[40]. Wang *et al* (2012, PNNL) also calculated that the radial peak of thermalization distance is approximately 30 nm for fluorides (CaF$_2$ and BaF$_2$) and 50 nm for iodides (NaI and CsI ). Further calculation from Prange *et al.*[18] (2017, PNNL) showed that the thermalization distance ($r_{hot}$) is larger and the fraction of free carriers ($\eta_{e,h}$) is smaller compared with the results of Wang *et al*. Both kMC simulation shows a sophisticated behaviour that $\eta_{e,h}$ varies with the incident γ-ray energy. Because lower incident energy means a higher ionization density and thus stronger electrostatic filed and higher percentage of electron-hole pairs that recombine during electron thermalization. Moreover, not only the thermalization time, the thermalization distance is also dependent on the group velocity of hot electrons. Calculations show that monovalent/simple alkali halides (NaI and CsI) tends to have a longer thermalization distance than multivalent/complex systems (YAP, SrI2, and BaBrI/Cl). For simplicity, the fraction of free carriers is considered as a constant and the fitted value $\eta_{e,h} = 18\%$ from Payne-Onsager model for LaBr$_3$:5%Ce was taken as a reference. A step-wise time-dependent electron diffusion coefficient such proposed by Lu et al. was adopted as well, so that D$_e$(t<τ$_{hot}$) has a constant value that reproduces the result of $r_{hot}$(peak) (<50 nm) in the solution of Eq. (1) at the end of τ$_{hot}$ ≈ 1 ps. Pairs of ($\eta_{e,h}, r_{hot}$), with $\eta_{e,h}$ from 0.1 to 0.2 to 0.5 and $r_{hot}$ from 10 nm to 25 nm to 50 nm, are simulated and compared with the experimental data, shown in Fig. 2. With this small $\eta_{e,h} = 18\%$, the non-proportionality curve is not sensitive to the thermalization distance, since most free carriers have combined each during thermalization process. A value of 25 nm, small than the 50 nm for CsI, is chosen for the multivalent LaBr$_3$. To conclude, higher $\eta_{e,h}$ and r$_{hot}$ show more "halide hump", which is in accordance with previous researches. LaBr$_3$:Ce belongs to multivalent and possesses a flat electron response and very small "halide hump".

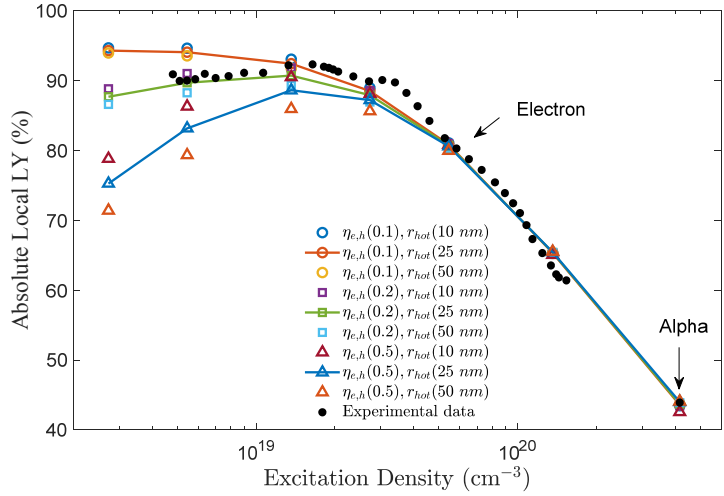

Fig. 2. Simulations of the absolute local light yield with excitation density under different thermalization parameters, along with the measured experimental data.

The mean thermalization time $\tau_{hot} \approx 1$ ps, due to its very similar $\omega_{LO}$ with NaI crystal, was further confirmed by the picosecond absorption spectroscopy experiment to be less than 1 ps under two-photon excitation of the host producing carriers near the band edges.

$K_{1e}$ and $K_{1h}$ are the electron and hole trapping rate of deep defects. The capture of deep defects is the cause of afterglow, and thus a large $K_{1e} = 2.7 \times 10^{10}\ s^{-1}$ was fitted in the CsI crystal. Whereas $K_{1h}$ is negligible compared with $K_{1e}$ due to the much smaller velocity approaching the trap in alkali halide[13]. However, the residual intensity induced by the X-ray exposure of LaBr$_3$:Ce was measured to be less than 0.01% after 200 s[47]. First principle calculation from Aberg *et al*. shows that in the ground state of LaBr$_3$:Ce, Ce preferentially substitutes for La with small distortions and adopts a neutral charge state corresponding to a Ce$^{3+}$-4f$^1$5d$^0$ configuration. The intrinsic defects of pure LaBr$_3$ is several orders smaller than LaBr$_3$ doped with 200 ppm Sr, which gives rise to a shallow acceptor substituting on a lanthanum site and thus improve the linearity of the photon light yield with respect to the energy of incident electron or photon. Moreover, the efficiency of light emission for LaBr$_3$:Ce is sufficiently high, so that free carriers are merely trapped by deep defects. A very small number of $K_{1e} = 1 \times 10^8\ s^{-1}$, not much diference to the value $K_{1e} = 0\ s^{-1}$, was used to kill the out-diffused and stopped free electrons and would not be driven back to form STE with STH for low ionization density. We also estimate the influence of $K_{1e}$ by changing its value from $0 \sim 1 \times 10^{10}\ s^{-1}$, the light yield of low excitation density changes less than 6%, while it has little influence on the high excitation density due to the competition process of recombination. $K_{1h}$ is neglected as 0, due to the small intrinsic deep defect and much lower velocity compared with electron.

$K_3$ is the third-order Auger recombination rate of free carriers. We evaluated $K_3$ from 0 to $6.6 \times 10^{-29}\ cm^6 s^{-1}$ and nothing changed. We found that $K_3$ does not affect the nonlinear quenching, since the hot diffusion of free electrons rapidly reduces the ionisation density.

The energy transfer rate $S_T$, from STEs to $Ce^{3+}$ ions, and the decay rate $R_{Ce3+*}$ of the excited $Ce^{3+*}$ can be measured precisely from the decay time profile of the gamma rays with the TCSPC method. The thermalization process, which lasts several picoseconds, rarely influences the decay profile with the time scale of nanoseconds. Moreover, the higher-order quenching terms can also be neglected for low-ionization density gamma rays, and thus will not influence the decay profiles. Detailed scintillation rise-time measurements in LaBr$_3$:Ce with fast coincidence methods by Glodo *et al.* [35] and Seifert *et al.* [31] have identified a fast stage ($S_{fast}$) and a slower stage ($S_{slow}$) of scintillation rise reported as approximately ~300 ps and ~2.1 ns, respectively. The percentage of the fast process ($A_{fast}$) is approximately 72%. The 300-ps process is further suggested by Li *et al.*[38] as the dipole-dipole transfer to Ce from STEs created in the neighborhood, and the 2.1-ns process is the thermally activated migration and energy transfer at room temperature in 5%Ce-doped LaBr3.

The dipole-dipole quenching parameters of $K_{2Ce*}$ and $K_{2E}$, mainly decrease the light yield for particles with high ionization densities, were calculated from the α/β ratio of internal radioactive alpha particles of CeBr$_3$ and LaBr$_3$:5%, respectively.

The conservation and thermalization stage normally takes place in the first several picoseconds, during which the nonlinear quenching is already completed or pre-determined. For halides, the competition between the outrun of hot electrons, thus captured by defects, and the formation of STEs within the inner static electric field strongly influence the quantity of carriers effective for further evolution. For alpha particles, of which the LET is higher than 300 MeV cm$^2$ g$^{-1}$, the independent fraction of free electrons and holes is near zero and STEs are formed effectively under a strong electrostatic force[48]. The model for alpha particles can be simplified as the one proposed by Bizarri and Dorenbos[36] but with the dipole-dipole quenching terms added for high ionization, in which STEs are formed instantly:

$$\frac{dN_{STE}}{dt} = G_{STE} - (R_{STE} + Q_{STE} + S_T) N_{STE} - K_{2E}(t) N_{STE}^2 \qquad (9)$$

$$\frac{dN_{Ce^*}}{dt} = S_T N_{STE} - R_{Ce^*} N_{Ce^*} - K_{2Ce^*}(t) N_{Ce^*}^2 \qquad (10)$$

$$S_T = S_T^0 \cdot \frac{n_{Ce}^0 - N_{Ce^*}}{n_{Ce}^0} \qquad (11)$$

The energy transfer rate of STEs and the decay rate of excited $Ce^{3+}$ ions have already been measured with gamma excitation and previously discussed. What remains unknown are the two second-order rate constants for the quenching terms of the STEs and the excited $Ce^{3+}$ ions. Fortunately, the second-order rate constants $K_{2Ce*}$ can be calculated using $CeBr_3$ crystals. The very fast transport process, 165-ps rise time $t_{10-90\%}$ for X-ray excitation, from free electrons/holes to the $Ce^{3+}$ ions is neglected, which may have limited variation of the calculated parameter $K_{2Ce*}$. The excitation of $CeBr_3$ can be considered as the direct excitation of $Ce^{3+}$ ions and the kinetic equation for $CeBr_3$ is then:

$$\frac{dN_{Ce^*}}{dt} = G_{Ce^*} - R_{Ce^*} N_{Ce^*} - K_{2Ce^*}(t) N_{Ce^*}^2 \tag{12}$$

The generation term $G_{Ce*}$ can be calculated using Eq.(6) with $\beta E_{gap}$=16.7 eV calculated from the light yield of 60000 Ph/MeV. The decay rate is the inverse of the measured decay time 17 ns. Using the local light yield for the alpha particles, the calculated $K_{2Ce*}$ is $0.097 \times 10^{-15}$ cm$^3$ s$^{-1/2}$. After the determination of $K_{2Ce*}$, the second-order rate constants $K_{2E}$ can be similarly ascertained using the local light yield for the alpha particles in LaBr$_3$:Ce by solving Eqs. (9)-(11). The calculated $K_{2E}$ is then $0.051 \times 10^{-15}$ cm$^3$ s$^{-1/2}$.

Table I. Parameters (and their literature references or comments on methods) used for the calculation of the light yield, proportionality, and scintillation time profile of LaBr$_3$:Ce at 295 K

| Parameter | Value | Units | Refs. and notes |
|---|---|---|---|
| $r_0$ | 3 | nm | Refs. [10,40] for NaI, Ref. [13] for CsI |
| $\beta E_{gap}$ | 13 ($\beta$=2.2) | (eV/e-h)$_{avg}$ | LY=77000 ph/MeV for LaBr$_3$:Ce,Sr[5] |
| $\varepsilon_0$ | 10 | N/A | Ref.[39,42] |
| $\eta_{e,h}$ | 0.18 | N/A | Payne-Onsager model, Ref. [11] |
| $\mu_e$ | 2 | cm$^2$ V$^{-1}$ s | Ref.[42] |
| $D_e(t > t_{hot})$ | 5.1×10$^{-2}$ | cm$^2$ s$^{-1}$ | $D_e = \mu_e kT/e$ |
| $\tau_{hot}$ | 1 | ps | Ref.[10,42] for NaI (with same $\omega_{LO}$) |
| $\mu_h$ | 1×10$^{-4}$ | cm$^2$ V$^{-1}$ s | Ref. [13] for CsI |
| $D_h$ | 2.6×10$^{-6}$ | cm$^2$ s$^{-1}$ | $D_h = \mu_h kT/e$ |
| $B(t > \tau_{hot})$ | 2.5×10$^{-7}$ | cm$^3$ s$^{-1}$ | Ref. [13,38] |
| $K_3$ | 4.5×10$^{-29}$ | cm$^6$ s$^{-1}$ | Ref.[13] for CsI, not sensitive |
| $K_{1e}$ | 1×10$^8$ | s$^{-1}$ | Ref. [13] |
| $K_{1h}$ | 0 | s$^{-1}$ | Ref [13,47] |
| $r_{hot}$ | 25 | nm | LaBr$_3$:5%Ce nPR curve, smaller than 50 nm. Ref. [49] |
| $D_e(t < t_{hot})$ | 3.1 | cm$^2$ s$^{-1}$ | to reproduce $r_{hot}$ at $\tau_{hot}$ |
| $R_{STE}$ | 1×10$^6$ | s$^{-1}$ | 1 us decay time of STE at 80 K, Ref.[47] |

| $Q_{STE}$ | 0 | s$^{-1}$ | negligible at room temperature. Ref.[36] |
| --- | --- | --- | --- |
| $S_{fast}$ | $1/(0.27\times10^{-9})$ | s$^{-1}$ | Ref. [31] |
| $S_{slow}$ | $1/(2.0\times10^{-9})$ | s$^{-1}$ | Ref. [31] |
| $A_{fast}$ | | N/A | Ref. [31] |
| $R_{Ce}$ | $1/(15.4\times10^{-9})$ | s$^{-1}$ | Ref. [31] |
| $D_{STE}$ | $2.6\times10^{-6}$ | cm$^2$ s$^{-1}$ | $D_{STE}\approx D_{STH}$, Ref. [50] |
| $K_{2Ce^*}$ | $9.7\times10^{-17}$ | t$^{-1/2}$ cm$^3$ s$^{-1/2}$ | Calculated from quenching factor of α particles with CeBr$_3$, Ref. [51] |
| $K_{2E}$ | $5.1\times10^{-17}$ | t$^{-1/2}$ cm$^3$ s$^{-1/2}$ | Calculated from quenching factor of α particles with LaBr$_3$:Ce, Ref. [51] |

## III. PROPORTIONALITY AND ITS IONIZATION DENSITY DEPENDENCE

### A. Experimental data

The intrinsic response of a scintillator is usually not proportional to the incident particle energy, which is normally considered caused by the nonlinear quenching of high ionization density as the electrons slow along the track. The light yield produced by internally generated electrons over a wide range of energies can be measured by the Compton coincidence [7,8,52,53] and Kdip [9,54] methods. Fig. 3(a). shows the measured electron responses of the LaBr$_3$:5%Ce$^{3+}$ crystal using the SLYNCI [41] and Kdip [33] methods at room temperature.

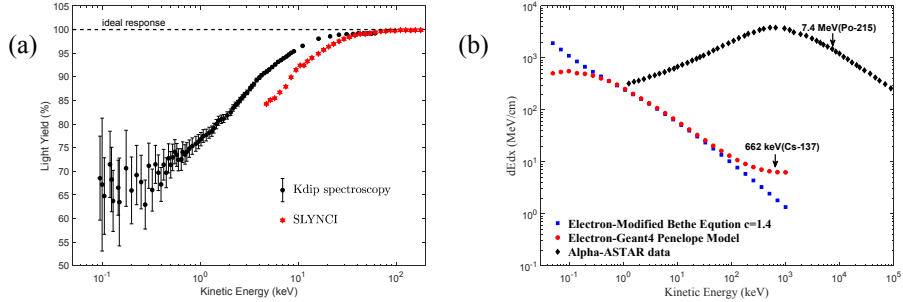

Fig. 3. (a) Combined plot of the two experiments for LaBr$_3$:5%Ce$^{3+}$ (300 K). The kinetic energy (keV) axis represents the electron energy. The light yield (%) axis represents the normalized light yield. The black dotted points are measured with the Kdip method and are available in [33], while the red hexagonal points are measured with the Compton coincidence method and are available in [41]. (b) dE/dx of the electrons with electron energy in LaBr$_3$:Ce.

The initial ionization density values vary from cell to cell along the length of the track with the variation in dE/dx and we calculate the local light yield for each local dE/dx value. The Penelope model from the Geant4 simulation is chosen to describe the variation in dE/dx with the electron energy. Although the modified Bethe equation was used by many previous researches [2,11,55-57], the Penelope model, shown in Fig. 3(b), can more properly describe the scintillation response of low-energy electrons, which tends to be constant between 0.1 keV and 0.4 keV.

Unlike the clustered track of CsI [13], the track of LaBr$_3$ is linear [58] so the intermediate local light yield can be calculated from experimental data. For an electron with an initial energy $E_0$, the measured light yield, $LY(E_0)$, is the average of the integration of the local light yield along the electron track. We then calculate the local light yield, $dL/dE(\varepsilon)$, for one cell with electron energy $\varepsilon$ using the following equations:

$$LY(E_0) = \int_0^{E_0} \frac{dL}{dE}(\varepsilon) d\varepsilon / E_0 \qquad (13)$$

$$\frac{dL}{dE}(\varepsilon) = \frac{d[\varepsilon \cdot LY(\varepsilon)]}{d\varepsilon} \qquad (14)$$

Similarly, the quenching of the alpha particles, which caused the nonlinear quenching in the high-density part of the ionization tracks, is also discussed [55]. The α/β ratio (or quenching factor) of the alpha particle and the corresponding calculated local light yield in the LaBr$_3$:Ce$^{3+}$ detector are shown in Table. II. Assuming an alpha particle with an initial energy of 7386.1 keV and decreasing its energy to 6819.2 keV, the stopping power from the ASTAR databases [55] of the particle changes little and can be considered a constant and represented by the mean values 301 MeV cm$^2$ g$^{-1}$. The mean local light yield, dL/dE, for an alpha particle with dE/dx of 1514 MeV cm$^{-1}$ (= 301 MeV cm$^2$ g$^{-1}$ × 5.03 g cm$^{-3}$) is 0.483.

Table II. The α/β ratio measured in different materials using time-amplitude analysis to separate the Po-215 and Rn-219 alpha peaks from the α internal background [51] and the calculated mean dL/dE of alpha particles with energy from 7386.1 keV to 6819.2 keV. The density [5] of LaBr$_3$:Ce is 5.03 g cm$^{-3}$ and CeBr$_3$ is 5.18 g cm$^{-3}$, which is used to convert the stopping power of the materials to dE/dx.

| Material | Isotope | True α energy (keV) | α/β ratio | Initial stopping power (MeV cm² g⁻¹) | Mean dE/dx (MeV cm⁻¹) | Mean dL/dE |
|---|---|---|---|---|---|---|
| LaBr$_3$:5%Ce | Po-215 | 7386.1 | 0.363 | 294 | 1514 | 0.483 |
| LaBr$_3$:5%Ce | Rn-219 | 6819.2 | 0.353 | 308 | | |
| CeBr$_3$ | Po-215 | 7386.1 | 0.266 | 294 | 1559 | 0.374 |
| CeBr$_3$ | Rn-219 | 6819.2 | 0.257 | 308 | | |

To conclude, the relationship between the local light yield and dE/dx for both the electron and alpha particles are shown in Fig. 4. They can be used to compare with our model directly with the change from dE/dx to excitation density with Eq.(6).

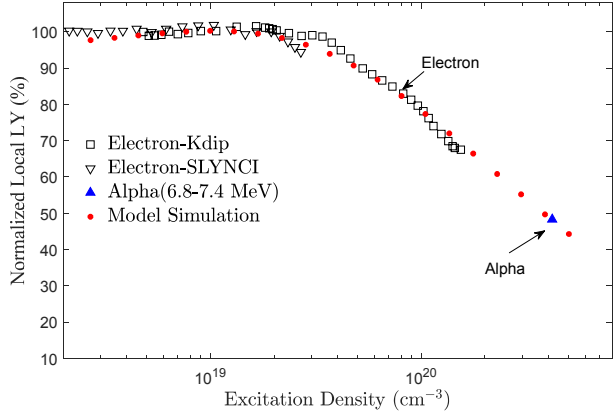

Fig. 4. Calculated normalized local light yield with the excitation density of LaBr$_3$:Ce from different experiments. The black open inverted triangles are the results measured with SLYNCI method, the black open squares are the resluts measured with Kdip method, the blue solid triangle is the data measured with 6.9-7.4 MeV alpha particles. The red dots are the simulation results of the model.

## B. Model results

Figure 4 shows the calculated normalized local light yield (normalized with excitation density of $2.7 \times 10^{18}$ cm$^{-3}$, corresponding to the LET of 10 MeV cm$^{-1}$) with the on-axis excitation density using the parameters of Table I in the Eqs. (1)-(5) for LaBr$_3$:5%Ce at room temperature. The red dots of our model fit the local light yield of both the electron and alpha particles very well. The simulated absolute light yield of 69.3 photons/keV for the low excitation density of the high-energy gamma rays is approximately 90% of the ideal LY, which corresponds to the measured light yield of 70 photons/keV. The ideal light yield of 77 photons/keV for LaBr$_3$:Ce was assumed as the best-known value of LaBr$_3$:Ce,Sr. The increase in the light yield for the co-doped LaBr$_3$:Ce,Sr was attributed to the efficient shallow trapper of the Sr defects and reduced the loss of the free electrons and holes by diffusing outward and being captured by the defects [59].

Since LaBr$_3$:Ce has linear track, the measured light yield, LY(E$_0$), is the average of the integration of the local light yield along the electron tracks using Eq.(13). The comparison between integrated light yield of the model and the experimentally measured light yield of the electrons with Kdip method is shown in Fig. 5, which shows good accordance for electrons with energy of 0.1-100 keV.

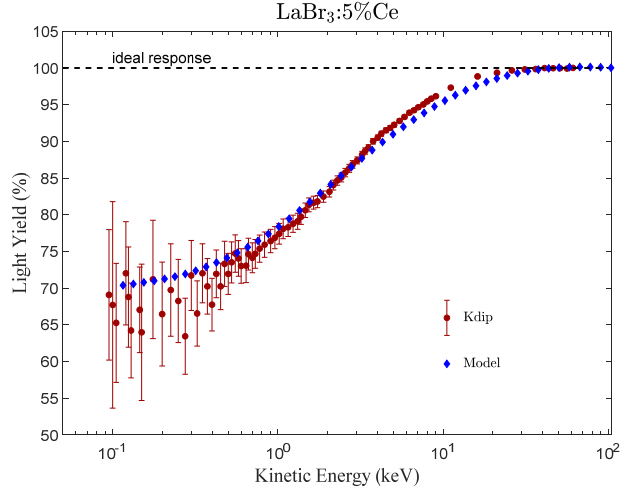

Fig. 5. The comparison between integrated light yield of the model and the measured light yield of the electrons.

## IV. PULSE SHAPE AND ITS IONIZATION DENSITY DEPENDENCE

The model includes the processes of linear energy transfer and radiative recombination, as well as nonlinear quenching, which will influence the excitation-density dependent pulse shapes. In this part, the decay curve of alpha/gamma can be calculated with the model and will be verified with experiments in the following sections.

### A. Experimental setup and experimental data

A cylindrical $LaBr_3$:5%$Ce^{3+}$ crystal was used. Since $LaBr_3$:$Ce^{3+}$ is hygroscopic, the sample was packaged in a metal can with a quartz window to prevent long-term exposure to moisture. The scintillation time profiles under excitation of both α particles from intrinsic $^{227}$Ac contamination and γ rays from a $^{22}$Na source were measured using the delayed coincidence method [60], shown in Fig. 6.

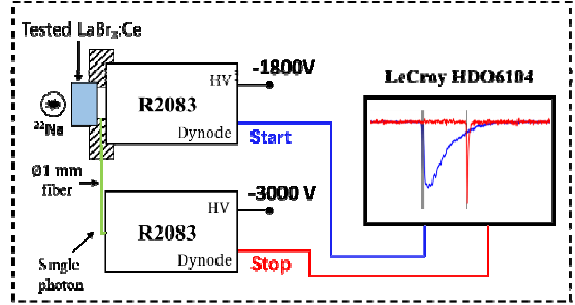

Fig. 6. Setup of time-correlated single photon counting measurement

Two R2083 PMTs were used as the start and stop detectors. A LeCroy HDO6104 oscilloscope (2.5 Gsps, 12 bit) was used to digitise the pulse shape of the start detector and a single photon signal of the stop detector to extract the time difference between the start and stop signal. The high voltage (HV) supply of the start detector was 1800 V due to the saturation of the start detector, while the HV for the stop detector was 3000 V to achieve better timing resolution.

In our experiment, the start channel collected approximately one-third of the total light output. The peak-over-charge ratio ($V_p/Q_{total}$, which is the area normalized amplitude, also known as A/E) [23] with the charge are shown in Fig. 7(a). The pulse shape discrimination (PSD) feature A/E was used to choose alpha events from an environmental gamma background, which was similar to the CCM PSD feature used in our previous work [25]. The averaged pulse shapes, each with 2000 events, of the two coloured rectangle regions are aligned by the 20% fraction of the peak amplitude and shown in Fig. 7(b). The averaged pulse shape of the alpha events was faster than the gamma events both in the rising part and the decay part.

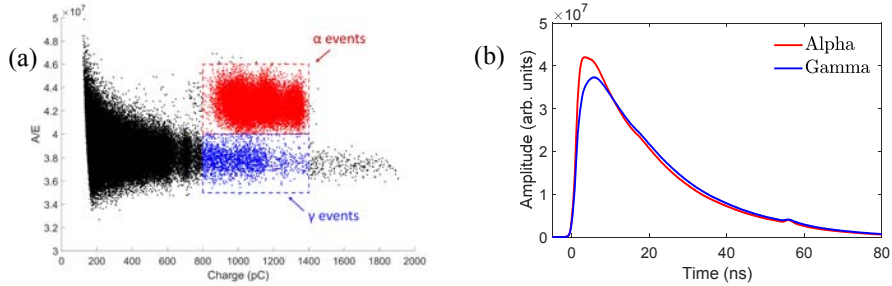

Fig. 7. (a) The distribution of the PSD feature A/E with energy. (b) Averaged pulse shapes of the events within the two coloured rectangle regions.

The measured time profile of both the α and γ particles (using a $^{22}$Na source) with the TCSPC method is shown in Fig. 8. Due to the relatively small volume (~0.393 cm$^3$) of crystals and the low alpha background count rate (0.082 cps cm$^{-3}$), the 30-day experimental data and 34382 effective alpha events (single photon signal) were collected for the TCSPC measurement.

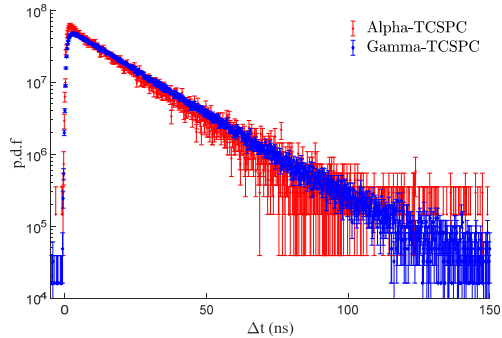

Fig. 8. Normalized time profiles of 511-keV gamma rays and alpha particles in the LaBr$_3$:Ce$^{3+}$ detector, measured with TCSPC method.

## B. System response of TCSPC experimental setup

Before the verification of the model, the *impulse response function* (IRF) of the measured time profiles, which mainly consists of two factors, need to be determined. First, the timing resolution of the start and stop detectors will convolute approximately a Gaussian response. Second, the photon transportation and bulk reabsorption following with the reemission due to the finite volume of the scintillation crystal also influence the scintillating time profiles characterized by the arrival times of collected photons.

1）characterization of the timing resolution for TCSPC measurement

Two identical LYSO crystals and a $^{22}$Na source were used to calibrate the time resolution of the start detector (LaBr$_3$:Ce$^{3+}$), shown in Fig. 9.

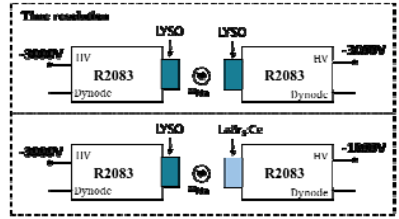

Fig. 9. Setup of calibration of TCPSC system time resolution

The overall Gaussian IRF function of the system is then described by the convolution of the two Gaussian functions with a total FWHM of:

$$\sqrt{(214\ \text{ps})^2 + (370\ \text{ps})^2} = 427\ \text{ps}$$

2) Characterization of photon transportation and bulk reabsorption with Monto Carlo simulation

A Monto Carlo simulation of the photon collection of scintillator, including photon transportation and bulk reabsorption, based on Geant4 was carried out. For the simulation, photons are generated uniformly within the scintillator and collected by the optical coupling shown in Fig. 10(a). The parameters of the energy distribution, scattering length, and absorption length of the generated photons were taken from the experiment carried out by Herman, et al. [61]. The reemission of absorbed photons is not considered in the simulation. One million photons were generated uniformly and 85.3% of the photons were collected without absorption. The distribution of the photons' arrival times, which is not negligible with the mean transportation time of 372 ps compared with the fast rising component of 300 ps, is shown in Fig. 10(b).

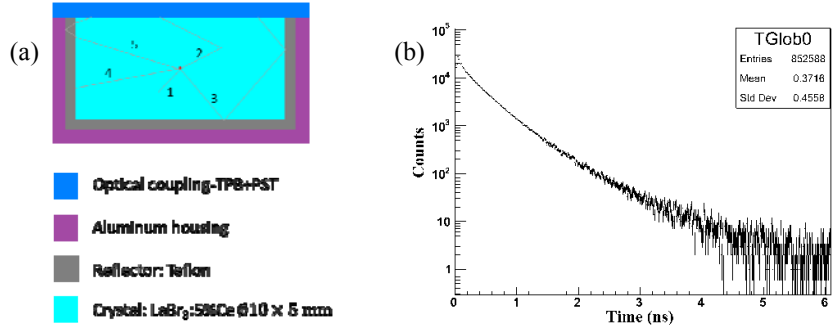

Fig. 10. (a) Illustration of the Monto Carlo simulation. (b) Distribution of the photons' arrival times within a ⌀10×5 mm crystal.

A modified reabsorption model from the one in SrI$_2$:Eu [62] was used,

$$I(t) = \sum_i p_i I_i(t) = \frac{1}{\tau} \sum_i p_i \left(\frac{t}{\tau}\right)^{i-1} \frac{\exp(-t/\tau)}{(i-1)!}, \tag{15}$$

where $I(t)$ is the intensity of the emission in time $t$ and $p_i$ is the probability of the collected photons with $(i-1)$ reabsorptions. However, Eq. (15) neglects the photon transportation response and only convolutes the time delay caused by the reabsorption. Multiple transportations caused by multiple reabsorptions are also considered in our model, which is especially not negligible for large crystals. The probability of the reabsorption of the $i$th remitted photon was considered the same, which was (1 - 85.3%) in our case, and neglected the possible change caused by a small shift in the emission that reduces the overlap between the emission and absorption [63]. The calculated response of photon collection, named as PHC, of LaBr$_3$:5%Ce is shown in Fig. 11 :

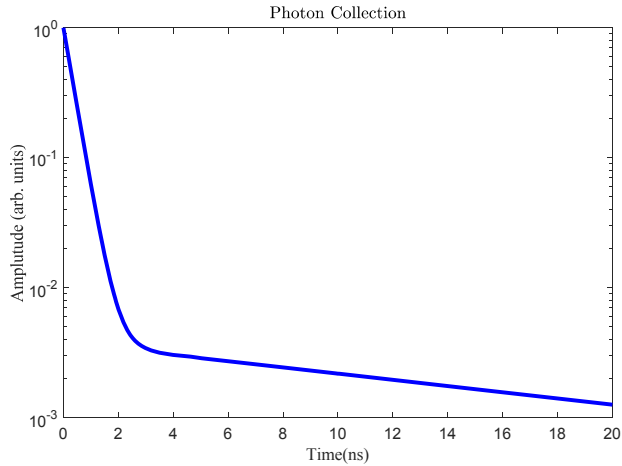

Fig. 11. Calculated response of photon collection, including multiple photon transportations and bulk reabsorptions.

## C. Model results

### 1）Verification of gamma pulse shape

The experimental data and model calculation, which convoluted the IRF of the system and the transportation and reabsorption of the crystal, are compared in Fig. 12. Superimposed is the original decay profile of the 511-keV gamma rays fitted by Seifert et al. [31] on which the energy transfer rates from the STE to $Ce^{3+}$ in our model were based. We tested our model results with the experimental data using the $\chi^2/ndf$ statistic:

$$\chi^2/ndf = \frac{1}{N}\sum_{n=1}^{N}\frac{(M_n - E_n)^2}{E_n}, \qquad (16)$$

where $N$ is the total number bins and $M_n$ is the measured number of entries in the nth histogram bin. $E_n$ is the expected number of counts in the nth bin based on our model.

The calculated $\chi^2/ndf$ between the experimental data and our model is 1.223 for gamma rays, which shows a good correspondence between our model and the experiment. It also suggests that the rate parameters describing the pulse shape of the gamma rays in our $LaBr_3:Ce^{3+}$ crystal is similar to those used in Ref. [31].

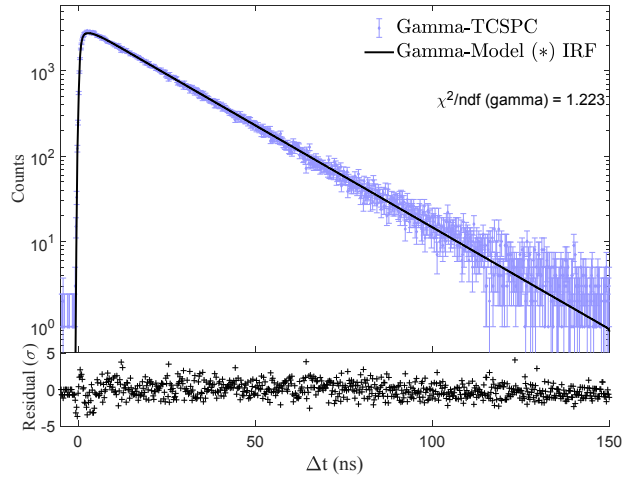

Fig. 12. The light blue is the measured decay profile of the 511 keV gamma rays. The dotted black line is the model results with the convolution of the system IRF, photon transportation, and reabsorption. The red line is the original decay profile of the 511 keV gamma rays fitted by Seifert et al. [31] measured with a small crystal.

2）**verification of alpha pulse shape**

With the determination of all of the parameters from the gamma-ray pulse shape and the quenching of the alpha particles, the calculated time response of the alpha particles can be assessed, with a mean stopping power of 440 MeV $cm^2$ $g^{-1}$ along the track. The comparison between the experimental data and model calculation after considering the system IRF and reabsorption is shown in Fig. 13(a), with $\chi^2/ndf = 1.011$, which demonstrates a good prediction of the scintillation time profile for alpha particles with high ionisation density. The detailed rising part between the measured time profiles and the model results is shown in Fig. 13(b).

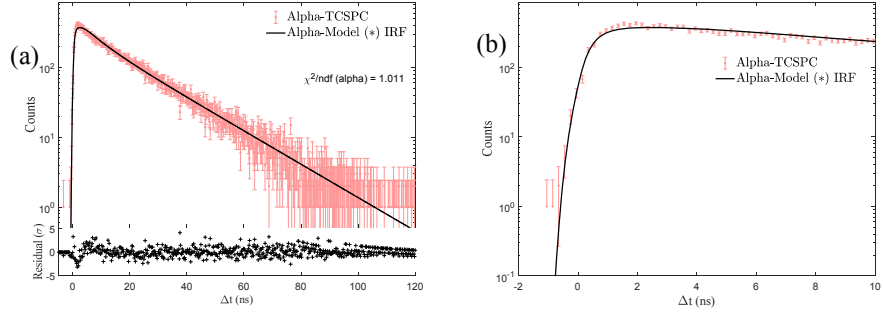

Fig. 13(a). The light red data is the measured decay profile of the alpha particles. The dotted black line is the model results of the alpha particles with the convolution of the system IRF, photon transportation, and reabsorption. (b) Detailed comparison of the rising part between the model calculation and the experimental data.

A summary of α/γ pulse shapes from experiment and model calculation is shown in Fig. 14.

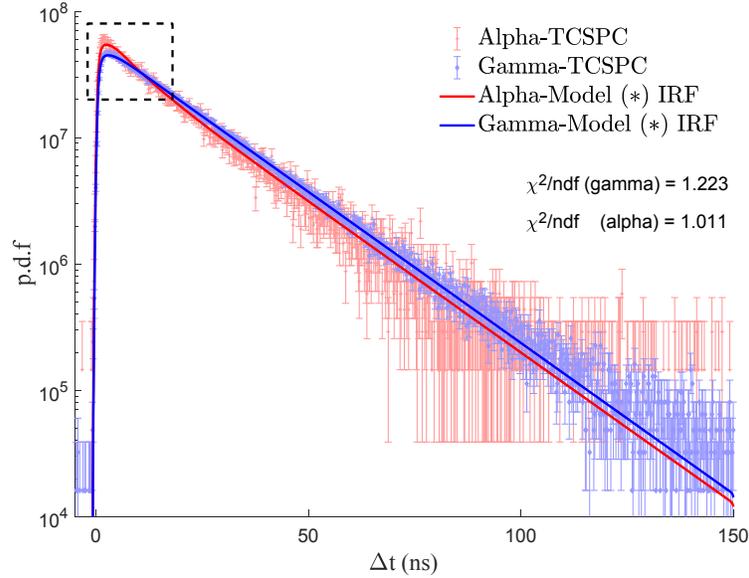

Fig. 14. Summary of α/γ pulse shapes from experiments and model calculations. The blue ones are for the gamma rays, while the red ones are for the alpha particles. The dots are for the measured decay curves with TCPSC method, while the solid curves are the calculated model results.

## V. DISCUSSON OF THE GENERALITY OF THE MODEL
### A. The α/γ ratio with the $Ce^{3+}$ concentration

According to research [55], the α/β ratio of $LaBr_3$:$Ce^{3+}$ changes with the Ce concentration and shows a parabolic shape with a maximum at 5% Ce dopant. A better α/β ratio means less quenching and better non-proportionality and thus good energy resolution. Herein we discuss the underlying cause of this phenomenon. As discussed in our previous model, two quenching processes of both STEs and $Ce^{3+*}$ can happen. Meanwhile, the rise time of the decay profile decreases with higher Ce concentrations due to the more efficient energy transfer process from STEs to $Ce^{3+}$ ions.

For lower Ce concentrations, less than 5% $Ce^{3+}$, the lifetime of the STEs becomes longer. Therefore, the quenching between the STEs plays a leading role and decreases the alpha/beta ratio. For $Ce^{3+}$ concentrations higher than 5%, with a $Ce^{3+}$ ion density $1.05 \times 10^{20}$ cm$^{-3}$, the quenching between the excited $Ce^{3+*}$ ions dominates and degrade the α/β ratio. The trade-off between these two quenching processes can qualitatively explain the shape of the α/β ratio and the best performance of 5% Ce concentration. To be more quantitatively, the α/β ratios were calculated using the proposed model of Eqs. (1)-(5) with only two parameter adjustments. The energy transfer rate, $S_T$, has been measured to be proportional to the Ce concentration. A linear extrapolation of the energy transfer rate was applied based on the 5% Ce situation. Moreover, with the increase of Ce concentration, the probability of the direct excitation of Ce dopant increases as well. The instant excited $Ce^{3+}$ ions, $G_{Ce^*}$, is assumed the same as the Ce concentration.

The α/β ratio of LaBr$_3$:Ce as a function of cerium concentration, measured with internal $^{215}$Po contamination, Eα = 7386 keV is shown as the black square in Fig. 15. The simulated α/β ratios of a particle with the ionization density of (560 × 5.1) MeV cm$^{-1}$ are also shown as the red triangles, which shows a reasonably good coincidence with the measurement. A further improvement of the model and the parameters can be carried out in the further with more experimental measurements.

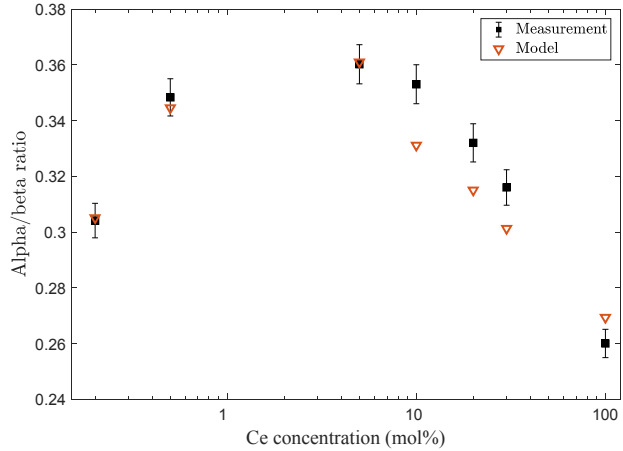

Fig. 15. Comparison of the α/β ratio of LaBr$_3$:Ce between simulation and measurement. The black rectangle points are the α/β ratio of LaBr$_3$:Ce as a function of cerium concentration , measured with internal 215Po contamination, Eα = 7386 keV. The red triangles are the. simulated α/β ratios of a particle with the ionization density of (560 × 5.1) MeV cm$^{-1}$.

## B. Dopant quenching and the pulse shape difference of LYSO and CeBr$_3$

LYSO:Ce and CeBr$_3$ are both fast scintillators and normally considered to possess only single decay component. The 10%-90% rise time of decay profile is 180 ps and 165 ps for LYSO and CeBr$_3$, respectively[64], which is associated with a certain energy transfer time constant of approximately 80 ps. The average pulse shapes of LYSO, from the direct output of R2083 PMT at 50 Ohm, under excitations of external $^{241}$Am-alpha and $^{22}$Na-gamma sources are shown in Fig. 16(a). Similarly, the average pulse shapes of CeBr$_3$ measured with internal low-activity $^{227}$Ac alpha contamination and external $^{22}$Na gamma source, are shown in Fig. 16(b).

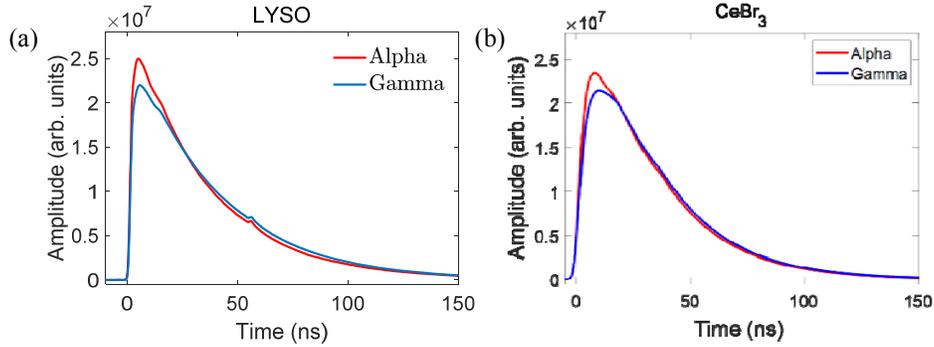

Fig. 16.(a) The averaged pulse shapes of a 5 mm LYSO cube under the excitation of a $^{241}$Am source. The blue line is the averaged pulse of the 59.5 keV gamma rays, and the red line is the averaged pulse of the alpha particle. (b) The averaged pulse shapes of a ⌀51×51 mm CeBr3 cylinder. The blue line is the averaged pulse of the 1500-2000 keV gamma rays, and the red line is the averaged pulse of the alpha particle.

Due to the very fast rising time of pulse shapes of LYSO and CeBr$_3$, the α/γ pulse shape difference could not be from the very fast energy transfer process. The faster decay profiles of the alpha particles indicate that the quenching between the luminescence centers, probably dopant Ce$^{3+}$ ions, can lead to the pulse shape differences of such fast scintillators. A detailed and quantitative study of LYSO and CeBr$_3$ on both non-proportionality and the pulse shape will be carried out in the future.

## VI. CONCLUSION

It has been confirmed that inorganic scintillators with single decay components also possess the ability of pulse shape discrimination between alpha and gamma particles, such as LaBr$_3$:Ce$^{3+}$ and LYSO, while the reason for the pulse shape difference in such single decay component scintillators remained unclear. In this paper, coupled rate and transport equations are established, taking typical single decay component scintillator LaBr$_3$:Ce$^{3+}$ as an example, to model the whole scintillation processes based on the previous studies on the scintillation mechanism of inorganic scintillators. With one parameter set, most of which can be measured with experiments and simulations, multiple observables of LaBr$_3$:Ce scintillation responses can be reproduced and further explained by the model, including the ionization density-dependent pulse shape differences for alpha or gamma particles, the proportionality response of electrons and quenching factor of alpha particles.

The following conclusions are drawn:

(1) With the quantitative discussion of the parameters and thus their influences on non-proportionality and pulse shape. The quenching process between the excited states of doped ions (Ce$^{3+*}$) is confirmed to mainly contribute observable ionization density-dependent α/γ pulse shape differences of single-decay component inorganic scintillator LaBr$_3$:Ce.

(2) The model reveals the relation between alpha quenching factor and Ce concentration in LaBr$_3$:Ce scintillator, which is in good accordance with previous experiments[55]. It is known that the less quenching of the light yield, the better proportionality and energy resolution the scintillators will have. Based on the model simulation, the quenching factor of LaBr$_3$:Ce achieves its maximum with 5% Ce concentration, which provides insight on the best energy resolution of 5% Ce concentration LaBr$_3$:Ce.

(3) Based on the study of the quenching process of excited excitation, the generality of ionization density-dependent pulse shape differences in other fast single-decay-component inorganic scintillators are predicted, and have been observed in LYSO and CeBr3 during our preliminary experiments.

Moreover, this model reveals the ionization density-dependent correlations between non-proportionality of light yield and scintillation pulse shape, which could provide a theoretical basis for the potential possibility of pulse by pulse correction of non-linear quenching with pulse shapes to achieve a better energy resolution, not only for LaBr$_3$:Ce. And we noticed such kind of correction were preliminarily explored in CsI:Tl[20,65] and NaI:Tl[65] scintillators recently. The establishment of a more quantitative and accurate model of physical processes will not only help understand the physical principles of scintillators, improve the crystal engineering, but may also reveal potential new methods of particle detection.

## Acknowledgements

This work was supported by the Tsinghua University Initiative Scientific Research Program

and the National Natural Science Foundation of China (Grant No. 11961141015).

# References:

[1] P. Dorenbos, J.T.M. De Haas, and C.W.E. van Eijk, Non-Proportionality in the Scintillation and the Energy Resolution with Scintillation Crystals, Ieee T Nucl Sci, **42**, 6, (1995).

[2] S.A. Payne, N.J. Cherepy, G. Hull, J.D. Valentine, W.W. Moses, and W. Choong, Nonproportionality of Scintillator Detectors: Theory and Experiment, Ieee T Nucl Sci, **56**, 2506, (2009).

[3] M. Moszyński, Energy resolution and non-proportionality of scintillation detectors – new observations, Radiat Meas, **45**, 372, (2010).

[4] I.V. Khodyuk and P. Dorenbos, Trends and Patterns of Scintillator Nonproportionality, Ieee T Nucl Sci, **59**, 3320, (2012).

[5] M. Nikl and A. Yoshikawa, Recent R&D Trends in Inorganic Single-Crystal Scintillator Materials for Radiation Detection, Adv Opt Mater, **3**, 463, (2015).

[6] M. Moszyński, A. Syntfeld-Każuch, L. Swiderski, M. Grodzicka, J. Iwanowska, and P. Sibczyński, Energy resolution of scintillation detectors, Nuclear Instruments and Methods in Physics Research Section A: Accelerators, Spectrometers, Detectors and Associated Equipment, **805**, 25, (2016).

[7] J.D. Valentine and B.D. Rooney, Design of a Compton spectrometer experiment for studying scintillator non-linearity and intrinsic energy resolution, Nuclear Instruments and Methods in Physics Research Section A: Accelerators, Spectrometers, Detectors and Associated Equipment, **353**, 37, (1994).

[8] W.S. Choong, K.M. Vetter, W.W. Moses, G. Hull, S.A. Payne, N.J. Cherepy, and J.D. Valentine, Design of a Facility for Measuring Scintillator Non-Proportionality, Ieee T Nucl Sci, **55**, 1753, (2008).

[9] I.V. Khodyuk, J.T.M. de Haas, and P. Dorenbos, Nonproportional response between 0.1-100 keV energy by means of highly monochromatic synchrotron X-rays, Ieee T Nucl Sci, **57**, 1175, (2010).

[10] J. Grim, K. Ucer, A. Burger, P. Bhattacharya, E. Tupitsyn, E. Rowe, V. Buliga, L. Trefilova, A. Gektin, and G. Bizarri, et al., Nonlinear quenching of densely excited states in wide-gap solids, Phys Rev B, **87**, (2013).

[11] S.A. Payne, W.W. Moses, S. Sheets, L. Ahle, N.J. Cherepy, B. Sturm, S. Dazeley, G. Bizarri, and W. Choong, Nonproportionality of Scintillator Detectors: Theory and Experiment. II, Ieee T Nucl Sci, **58**, 3392, (2011).

[12] Q. Li, J.Q. Grim, R.T. Williams, G.A. Bizarri, and W.W. Moses, The role of hole mobility in scintillator proportionality, Nuclear Instruments and Methods in Physics Research Section A: Accelerators, Spectrometers, Detectors and Associated Equipment, **652**, 288, (2011).

[13] X. Lu, Q. Li, G.A. Bizarri, K. Yang, M.R. Mayhugh, P.R. Menge, and R.T. Williams, Coupled rate and transport equations modeling proportionality of light yield in high-energy electron tracks: CsI at 295 K and 100 K; CsI:Tl at 295 K, Phys Rev B, **92**, 115207, (2015).

[14] X. Lu, S. Gridin, R.T. Williams, M.R. Mayhugh, A. Gektin, A. Syntfeld-Kazuch, L. Swiderski, and M. Moszynski, Energy-Dependent Scintillation Pulse Shape and Proportionality of Decay Components for CsI:Tl: Modeling with Transport and Rate Equations, Phys Rev Appl, **7**, 14007, (2017).

[15] S. Kerisit, K.M. Rosso, and B.D. Cannon, Kinetic Monte Carlo Model of Scintillation Mechanisms in CsI and CsI(Tl), Ieee T Nucl Sci, **55**, 1251, (2008).

[16] Z. Wang, Y. Xie, B.D. Cannon, L.W. Campbell, F. Gao, and S. Kerisit, Computer simulation of electron thermalization in CsI and CsI(Tl), J Appl Phys, **110**, 64903, (2011).

[17] D. Åberg, B. Sadigh, A. Schleife, and P. Erhart, Origin of resolution enhancement by co-doping of scintillators: Insight from electronic structure calculations, Appl Phys Lett, **104**, 211908, (2014).

[18] M.P. Prange, Y. Xie, L.W. Campbell, F. Gao, and S. Kerisit, Monte Carlo simulation of electron thermalization in scintillator materials Implications, J Appl Phys, **122**, 234504, (2017).

[19] A. Syntfeld-Kazuch, M.M. Ski, A. Widerski, W. Klamra, and A. Nassalski, Light Pulse Shape Dependence on $\gamma$-ray Energy in CsI, Ieee T Nucl Sci, **55**, 1246, (2008).

[20] S. Gridin, D.R. Onken, and R.T. Williams, Pulse Shape Analysis of Individual Gamma Events-Correlation to Energy Resolution and the Possibility of its Improvement, J. Appl. Phys, **154504**, (2018).

[21] C. Hoel, L.G. Sobotka, K.S. Shah, and J. Glodo, Pulse-shape discrimination of La halide scintillators, Nuclear Instruments and Methods in Physics Research Section A: Accelerators, Spectrometers, Detectors and Associated Equipment, **540**, 205, (2005).

[22] F.C.L. Crespi, F. Camera, N. Blasi, A. Bracco, S. Brambilla, B. Million, R. Nicolini, L. Pellegri, S. Riboldi, and M. Sassi, et al., Alpha–gamma discrimination by pulse shape in $LaBr_3$:Ce and $LaCl_3$:Ce, Nuclear Instruments and Methods in Physics Research Section A: Accelerators, Spectrometers, Detectors and Associated Equipment, **602**, 520, (2009).

[23] R. Ogawara and M. Ishikawa, Feasibility study on signal separation for spontaneous alpha decay in LaBr3:Ce scintillator by signal peak-to-charge discrimination, Rev Sci Instrum, **8**, 85108, (2015).

[24] J.E. McFee, C.M. Mosquera, and A.A. Faust, Comparison of model fitting and gated integration for pulse shape discrimination and spectral estimation of digitized lanthanum halide scintillator pulses, Nuclear Instruments and Methods in Physics Research A, **828**, 105, (2016).

[25] M. Zeng, J. Cang, Z. Zeng, X. Yue, J. Cheng, Y. Liu, H. Ma, and J. Li, Quantitative Analysis and Efficiency Study of PSD Methods for a $LaBr_3$:Ce Detector, Nuclear Instruments and Methods in Physics Research Section A: Accelerators, Spectrometers, Detectors and Associated Equipment, **813**, 56, (2016).

[26] M.P. Taggart and J. Henderson, Fast-neutron response of $LaBr_3$(Ce) and $LaCl_3$(Ce) scintillators, Nuclear Instruments and Methods in Physics Research Section A: Accelerators, Spectrometers, Detectors and Associated Equipment, **975**, 164201, (2020).

[27] H. Cheng, B. Sun, L. Zhu, T. Li, G. Li, C. Li, X. Wu, and Y. Zheng, Intrinsic background radiation of $LaBr_3$(Ce) detector via coincidence measurements and simulations, Nucl Sci Tech, **31**, 99, (2020).

[28] T. Alharbi, Pulse-shape discrimination of internal $\alpha$-contamination in LaBr3:Ce detectors by using the principal component analysis, J Instrum, **15**, P6010, (2020).

[29] K. Yang and P.R. Menge, Enhanced α-γ Discrimination in Co-doped LaBr$_3$:Ce, in *IEEE Nuclear Science Symposium & Medical Imaging Conference (NSS/MIC)* Seattle, 2014.

[30] K. Yang, P.R. Menge, and V. Ouspenski, Enhanced Discrimination in Co-doped LaBr$_3$:Ce, Ieee T Nucl Sci, **63**, 416, (2016).

[31] S. Seifert, J.H.L. Steenbergen, H.T.V. Dam, and D.R. Schaart, Accurate measurements of the rise and decay times of fast scintillators with solid state photon counters, J Instrum, **7**, P9004, (2012).

[32] Q. Li, J.Q. Grim, K.B. Ucer, A. Burger, G.A. Bizarri, W.W. Moses, and R.T. Williams, Host structure dependence of light yield and proportionality in scintillators in terms of hot and thermalized carrier transport, physica status solidi (RRL) - Rapid Research Letters, **6**, 346, (2012).

[33] W.W. Moses, G.A. Bizarri, R.T. Williams, S.A. Payne, A.N. Vasil'Ev, J. Singh, Q. Li, J.Q. Grim, and W. Choong, The Origins of Scintillator Non-Proportionality, Ieee T Nucl Sci, **59**, 2038, (2012).

[34] K.S. Shah, J. Glodo, M. Klugerman, W.W. Moses, S.E. Derenzo, and M.J. Weber, LaBr3:Ce scintillators for gamma-ray spectroscopy, Ieee T Nucl Sci, **50**, 2410, (2003).

[35] J. Glodo, W.W. Moses, W.M. Higgins, E.V.D. van Loef, P. Wong, S.E. Derenzo, M.J. Weber, and K.S. Shah, Effects of Ce concentration on scintillation properties of LaBr$_3$:Ce, Ieee T Nucl Sci, **52**, 1805, (2005).

[36] G. Bizarri and P. Dorenbos, Charge carrier and exciton dynamics in LaBr$_3$:Ce$^{3+}$ scintillators: Experiment and model, Phys Rev B, **75**, 184302, (2007).

[37] M.S. Alekhin, S. Weber, K.W. Krämer, and P. Dorenbos, Optical properties and defect structure of Sr$^{2+}$ co-doped LaBr$_3$:5%Ce scintillation crystals, J Lumin, **145**, 518, (2014).

[38] P. Li, S. Gridin, K.B. Ucer, and R.T. Williams, Picosecond absorption spectroscopy of self-trapped excitons and transient Ce states in LaBr$_3$ and LaBr$_3$:Ce, Phys Rev B, **97**, 144303, (2018).

[39] B. Liu, M. Gu, Z. Qi, X. Liu, S. Huang, and C. Ni, First-principles study of lattice dynamics and thermodynamic properties of LaCl$_3$ and LaBr$_3$, Phys Rev B, **76**, 64307, (2007).

[40] Z. Wang, Y. Xie, L.W. Campbell, F. Gao, and S. Kerisit, Monte Carlo simulations of electron thermalization in alkali iodide and alkaline-earth fluoride scintillators, J Appl Phys, **112**, 14906, (2012).

[41] R.M. Van Ginhoven, J.E. Jaffe, S. Kerisit, and K.M. Rosso, Trapping of Holes and Excitons in Scintillators: CsI and LaX$_3$ (X = Cl, Br), Ieee T Nucl Sci, **57**, 2303, (2010).

[42] I.V. Khodyuk, F.G.A. Quarati, M.S. Alekhin, and P. Dorenbos, Energy resolution and related charge carrier mobility in LaBr$_3$:Ce scintillators, J Appl Phys, **114**, 123510, (2013).

[43] Q. Li, J.Q. Grim, R.T. Williams, G.A. Bizarri, and W.W. Moses, A transport-based model of material trends in nonproportionality of scintillators, J Appl Phys, **109**, (2011).

[44] W. Zhiguo, W.R. T., G.J. Q., G. Fei, and K. Sebastien, Kinetic Monte Carlo simulations of excitation density dependent scintillation in CsI and CsI(Tl), physica status solidi (b), **250**, 1532, (2013).

[45] M.P. Prange, R.M. Van Ginhoven, N. Govind, and F. Gao, Formation, stability, and mobility of self-trapped excitations in NaI and NaI 1-x Tl x from first principles, Physical Review B - Condensed Matter and Materials Physics, **87**, (2013).

[46] P.R. Beck, S.A. Payne, S. Hunter, L. Ahle, N.J. Cherepy, and E.L. Swanberg, Nonproportionality of Scintillator Detectors. V. Comparing the Gamma and Electron Response, Ieee T Nucl Sci, **62**, 1429, (2015).

[47] G. Bizarri, J.T.M. de Haas, P. Dorenbos, and C.W.E. van Eijk, Scintillation properties of 1x1 Inch $LaBr_3$: 5%$Ce^{3+}$ crystal, Ieee T Nucl Sci, **53**, 615, (2006).

[48] R.T. Williams, J.Q. Grim, Q. Li, K.B. Ucer, G.A. Bizarri, and A. Burger, Scintillation Detectors of Radiation: Excitations at High Densities and Strong Gradients, in *Excitonic and Photonic Processes in Materials*, edited by J. Singh, and R.T. Williams (Springer, 2014), p. 299.

[49] J.Q. Grim, Q. Li, K.B. Ucer, A. Burger, G.A. Bizarri, W.W. Moses, and R.T. Williams, The roles of thermalized and hot carrier diffusion in determining light yield and proportionality of scintillators, Phys Status Solidi a, **209**, 2421, (2012).

[50] P. Erhart, A. Schleife, B. Sadigh, and D. Aberg, Quasiparticle spectra, absorption spectra, and excitonic properties of NaI and $SrI_2$ from many-body perturbation theory, Phys Rev B, **89**, (2014).

[51] W. Wolszczak and P. Dorenbos, Shape of intrinsic alpha pulse height spectra in lanthanide halide scintillators, Nuclear Instruments and Methods in Physics Research Section A: Accelerators, Spectrometers, Detectors and Associated Equipment, **857**, 66, (2017).

[52] B.D. Rooney and J.D. Valentine, Benchmarking the Compton Coincidence Technique for Measuring Electron Response Non-Proportionality in Inorganic Scintillators, Ieee T Nucl Sci, **43**, (1996).

[53] W. Choong, G. Hull, W.W. Moses, K.M. Vetter, S.A. Payne, N.J. Cherepy, and J.D. Valentine, Performance of a Facility for Measuring Scintillator Non-Proportionality, Ieee T Nucl Sci, **55**, 1073, (2008).

[54] I.V. Khodyuk, P.A. Rodnyi, and P. Dorenbos, Nonproportional scintillation response of NaI:Tl to low energy x-ray photons and electrons, J Appl Phys, **107**, 113513, (2010).

[55] W. Wolszczak and P. Dorenbos, Nonproportional Response of Scintillators to Alpha Particle Excitation, Ieee T Nucl Sci, **64**, 1580, (2017).

[56] A. Jablonski, S. Tanuma, and C.J. Powell, New universal expression for the electron stopping power for energies between 200 eV and 30 keV, Surf Interface Anal, **38**, 76, (2006).

[57] G. Bizarri, W.W. Moses, J. Singh, A.N. Vasil'Ev, and R.T. Williams, An analytical model of nonproportional scintillator light yield in terms of recombination rates, J Appl Phys, **105**, (2009).

[58] F. Gao, S. Kerisit, Y. Xie, D. Wu, M. Prange, R. Van Ginhoven, L. Campbell, and Z. Wang, Science-Driven Candidate Search for New Scintillator Materials FY 2013 Annual Report, (2013).

[59] M.S. Alekhin, J.T.M. de Haas, I.V. Khodyuk, K.W. Krämer, P.R. Menge, V. Ouspenski, and P. Dorenbos, Improvement of γ-ray energy resolution of $LaBr_3$:$Ce^{3+}$ scintillation detectors by $Sr^{2+}$ and $Ca^{2+}$ co-doping, Appl Phys Lett, **102**, 161915, (2013).

[60] L.M. Bollinger and G.E. Thomas, Measurement of the Time Dependence of Scintillation Intensity by a Delayed-Coincidence Method, Rev Sci Instrum, **32**, 1044, (1961).

[61] H.T. van Dam, S. Seifert, W. Drozdowski, P. Dorenbos, and D.R. Schaart, Optical Absorption Length, Scattering Length, and Refractive Index of $LaBr_3$:$Ce^{3+}$, Ieee T Nucl Sci, **59**, 656, (2012).

[62] J. Glodo, E.V. van Loef, N.J. Cherepy, S.A. Payne, and K.S. Shah, Concentration Effects in Eu Doped SrI2, Ieee T Nucl Sci, **57**, 1228, (2010).

[63] D.N. ter Weele, D.R. Schaart, and P. Dorenbos, The Effect of Self-Absorption on the Scintillation Properties of $Ce^{3+}$ Activated $LaBr_3$ and $CeBr_3$, Ieee T Nucl Sci, **61**, 683, (2014).

[64] D.N. ter Weele, D.R. Schaart, and P. Dorenbos, Intrinsic scintillation pulse shape measurements by means of picosecond x-ray excitation for fast timing applications, Nuclear Instruments and Methods in Physics Research Section A: Accelerators, Spectrometers, Detectors and Associated Equipment, **767**, 206, (2014).

[65] A. Gektin, A.N. Vasil'Ev, V. Suzdal, and A. Sobolev, Energy Resolution of Scintillators in Connection With Track Structure, Ieee T Nucl Sci, **67**, 880, (2020).